\documentclass[twocolumn,english,aps,prb,floats,showpacs]{revtex4}
\usepackage[T1]{fontenc}
\usepackage[latin9]{inputenc}
\usepackage{bm}
\usepackage{amsmath}
\usepackage{amssymb}
\usepackage{graphicx}
\usepackage{wasysym}
\usepackage{esint}

\makeatletter
\@ifundefined{textcolor}{}
{%
 \definecolor{BLACK}{gray}{0}
 \definecolor{WHITE}{gray}{1}
 \definecolor{RED}{rgb}{1,0,0}
 \definecolor{GREEN}{rgb}{0,1,0}
 \definecolor{BLUE}{rgb}{0,0,1}
 \definecolor{CYAN}{cmyk}{1,0,0,0}
 \definecolor{MAGENTA}{cmyk}{0,1,0,0}
 \definecolor{YELLOW}{cmyk}{0,0,1,0}
}

\@ifundefined{textcolor}{}{%
 \definecolor{BLACK}{gray}{0}
 \definecolor{WHITE}{gray}{1}
 \definecolor{RED}{rgb}{1,0,0}
 \definecolor{GREEN}{rgb}{0,1,0}
 \definecolor{BLUE}{rgb}{0,0,1}
 \definecolor{CYAN}{cmyk}{1,0,0,0}
 \definecolor{MAGENTA}{cmyk}{0,1,0,0}
 \definecolor{YELLOW}{cmyk}{0,0,1,0}
}

\usepackage{babel,hyperref,ifpdf}
\usepackage{bm}

\usepackage{babel}

\usepackage{babel}

\usepackage{babel}

\makeatother

\usepackage{babel}
\begin{document}

\title{Stabilty of Biskyrmions in Centrosymmetric Magnetic Films}

\author{Daniel Capic, Dmitry A. Garanin, and Eugene M. Chudnovsky}

\affiliation{Physics Department, Herbert H. Lehman College and Graduate School,
The City University of New York, 250 Bedford Park Boulevard West,
Bronx, New York 10468-1589, USA }

\date{\today}
\begin{abstract}
Motivated by the observation of biskyrmions in centrosymmetric magnetic
films (Yu et al. Nature Communications 2014, Wang et al. Advanced
Materials 2016), we investigate analytically and numerically the stability
of biskyrmions in films of finite thickness, taking into account the
nearest-neighbor exchange interaction, perpendicular magnetic anisotropy
(PMA), dipole-dipole interaction (DDI), and the discreteness of the
atomic lattice. The biskyrmion is characterized by the topological
charge $Q=2$, the spatial scale $\lambda$, and another independent
length $d$ that can be interpreted as a separation of two $Q=1$
skyrmions inside a $Q=2$ topological defect in the background of
uniform magnetization. We find that biskyrmions with $d$ of order
$\lambda$ can be stabilized by the magnetic field within a certain
range of the ratio of PMA to DDI in a film having a sufficient number
of atomic layers $N_{z}$. The shape of biskyrmions has been obtained
by the numerical minimization of the energy of interacting spins in
a $1000\times1000\times N_{z}$ atomic lattice. It is close to the
exact solution of Belavin-Polyakov model when $d$ is below the width
of the ferromagnetic domain wall. We compute the magnetic moment of
a biskyrmion and discuss ways of creating biskyrmions in experiment. 
\end{abstract}

\maketitle

\section{Introduction}

Magnetic skyrmions in 2D films represent a very active field of research
due to their potential for topologically protected data storage and
information processing at the nanoscale \cite{Nagaosa2013,Zhang2015,Klaui2016,Leonov-NJP2016,Hoffmann-PhysRep2017,Fert-Nature2017}.
Skyrmions were initially introduced in high energy physics as nonlinear
field models of elementary particles \cite{SkyrmePRC58,Polyakov-book,Manton-book}.
They entered condensed matter physics after it was realized that Skyrme's
theory described topological defects in ferro- and antiferromagnetic
films \cite{BelPolJETP75,Lectures,Brown-book}. Similar topology leads
to skyrmions in Bose-Einstein condensates \cite{AlkStoNat01}, quantum
Hall effect \cite{SonKarKivPRB93,StonePRB93}, anomalous Hall effect
\cite{YeKimPRL99}, and liquid crystals \cite{WriMerRMR89}.

Skyrmions are characterized by the topological charge $Q=\pm1,\pm2,...$.
In a $2D$ exchange model of a continuous spin field the conservation
of $Q$ is provided by topolgy: Different $Q$ arise from different
homotopy classes of the mapping of the three-component fixed-length
spin field onto the $2D$ plane. Similar topological properties are
possessed by the magnetic bubbles that have been intensively studied
in 1970s \cite{MS-bubbles,ODell}. They were in effect cylindrical
domains surrounded by domain walls of thickness $\delta$ that is
small compared to the radius of the domain $R$. In typical ferromagnets
$\delta\sim10$-$100$nm, so the bubbles of 1970's were at least of
a micron size or greater. With the emergence of nanoscience and nanoscale
measuring techniques, the experimentalists have been able to observe
topological defects in $2D$ magnetic films of size smaller than $\delta$.
This is when the field of ``skyrmionics'' took off in condensed
matter physics. Unlike the bubbles, nanoscale skyrmions are much closer
to the topological defects described by the Skyrme model.

\begin{figure}[ht]
\centering{}\includegraphics[width=8cm]{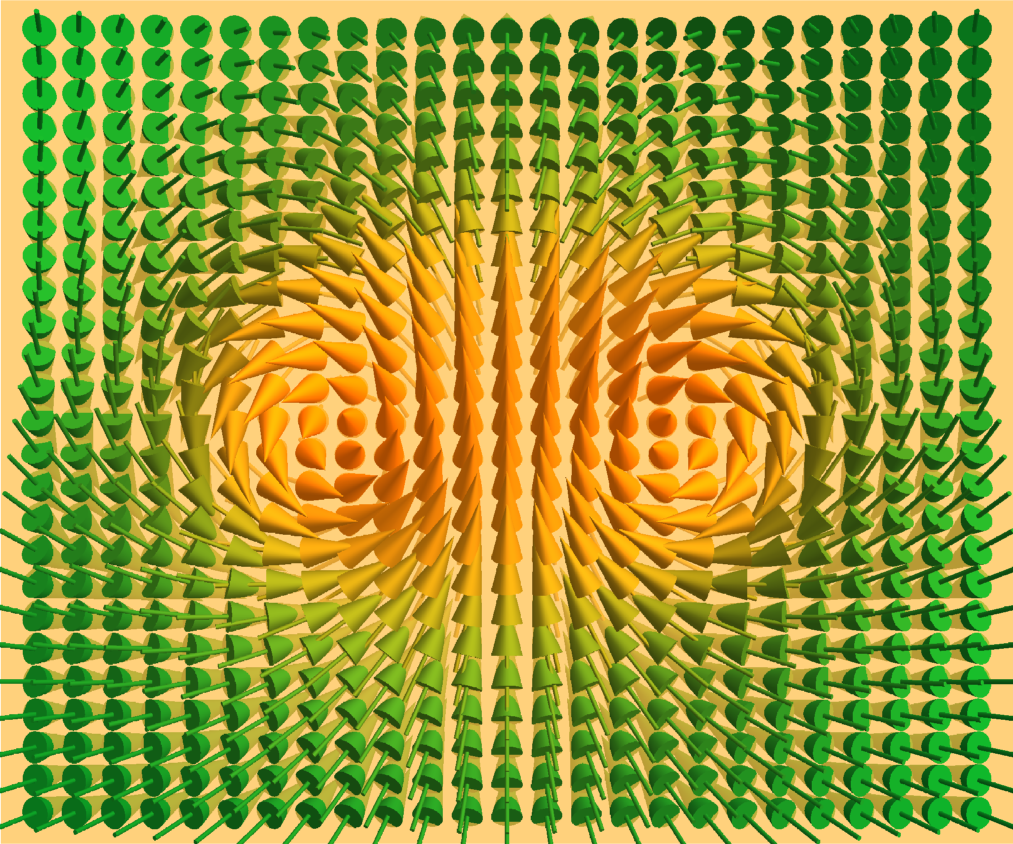} \caption{Computer generated spin field in a Bloch-type byskirmion in a ferromagnetic
film.}
\label{Fig-biskyrmion} 
\end{figure}

Research on magnetic skyrmions has focused on their stability, dynamics,
and various symmetry properties. Perpendicular magnetic anisotropy
(PMA), dipole-dipole interaction (DDI), magnetic field, and confined
geometry can stabilize significantly large magnetic bubbles \cite{IvanovPRB06,IvanovPRB09,Moutafis-PRB2009,Ezawa-PRL2010,Makhfudz-PRL2012}.
For small skyrmions, violation of the scale invariance by the crystal
lattice with a finite atomic spacing $a$ leads to a stronger violation
of the conservation of $Q$. In a pure exchange model, ferromagnetic
skyrmions of size $\lambda$ collapse \cite{CCG-PRB2012} on a time
scale proportional to $(\lambda/a)^{5}$. Their stability requires
other than Heisenberg exchange coupling, strong random field or random
anisotropy, or, most commonly, a non-centrosymmetric system with large
Dzyaloshinskii-Moriya interaction (DMI) \cite{AbanovPRB98,Bogdanov-Nature2006,Heinze-Nature2011,Leonov-NatCom2015,Chen-APL2015,Boulle-NatNano2016,Lin-PRB2016,Leonov-NJP2016,EC-DG-PRL2018,EC-DG-NJP2018}.

To date, the presence of stable biskyrmions (see the computer-generated
image in Fig.\ \ref{Fig-biskyrmion}) has been independently reported
at least in two centrosymmetric films of finite thickness: the La$_{2-2x}$Sr$_{1+2x}$Mn$_{2}$O$_{7}$
manganite \cite{Yu-2014} and the (Mn$_{1-x}$Ni$_{x}$)$_{65}$Ga$_{35}$
half Heusler alloy \cite{Zhang2016}. Given that stability of $Q=1$
skyrmions in films of large lateral dimensions normally require DMI,
these findings are quite amazing and call for a theoretical analysis.
Previously we have shown \cite{GCZ-EPL2017} that clusters with $Q>1$
are naturally generated due to the presence of Bloch lines in domain
walls when labyrinth domains are destroyed by the magnetic field in
a centrosymmetric magnetic film of finite thickness. We also observed
that $Q=2$ biskyrmions were generated by slow relaxation of the system
at $T=0$ starting from the disordered spin state or, equivalently,
by slow cooling to $T=0$ from high temperature. A more detailed numerical
investigation of biskyrmions, including the current-induced dynamics,
was performed in Ref. \onlinecite{XZhang2017} within a 2D frustrated
micromagnetic model. Biskyrmions arising from frustrated Heisenberg
exchange have also been reported in studies of triangular spin lattices
\cite{Leonov-NatCom2015} and in Ginzburg-Landau theory of skyrmions
\cite{Lin-PRB2016}. Metastable biskyrmion configurations have been
observed in Landau-Lifshitz dynamics of a frustrated bilayer film
\cite{Nowak2017}. However, the study of the separation of skyrmions
in a biskyrmion, the shape of the biskyrmion, and its stability on
the applied magnetic field, on the strength of the DDI, and on the
film thickness in centrosymmetric systems has been absent so far.
This, together with the above-mentioned experimental findings, provided
motivation for our work.

In the past, biskyrmions have been intensively studied in nuclear
physics in a hope that they would provide a model of a deuteron \cite{D1,D2}.
The Belavin-Polyakov (BP) pure exchange model \cite{BelPolJETP75}
in 2D contains an exact solution with $Q=2$ characterized by an arbitrary
spatial dimension $\lambda$, and another arbitrary parameter $d$
that can be visually interpreted as a separation of two $Q=1$ skyrmions
in a $Q=2$ topological defect (see Fig.\ \ref{Fig-biskyrmion}),
although the non-linearity of the model makes such interpretation
meaningful only at a large separation. At $d=0$ the defect possesses
symmetry with respect to the rotation in the $xy$-plane, which gets
broken by any $d\neq0$. We derive analytical formulas for the spin
components and the magnetic moment of a BP biskyrmion for arbitrary
$\lambda$ and $d$.

In real systems, the magnetic biskyrmions are more complicated as
they are formed by a number of competing interactions. In this paper
we show that a biskyrmion spin configuration naturally arises from
the energy minimization in a centrosymmetric film of finite thickness
in a lattice model that contains nearest-neighbor exchange interaction,
perpendicular magnetic anisotropy (PMA), and dipole-dipole interaction
(DDI). In accordance with the experimental findings, we find that
stable biskyrmions exist within a certain range of parameters in films
of sufficient thickness.

One interesting observation that follows from our studies is that
the DDI always favors a biskyrmion with a finite $d$ over a $Q=2$
topological defect that does not split into $Q=1$ skyrmions. Another
interesting observation is that, although biskyrmions are stabilized
by interactions other than ferromagnetic exchange, sufficiently small
biskyrmions are always close to the BP shape. On changing the parameters
and the external magnetic field one can change the shape of the $Q=2$
topological defect from a BP biskyrmion to a thin-wall biskyrmion
bubble.

This paper is organized as follows. Analytical formulas for the spin
components, the magnetic moment of a $Q=2$ skyrmion in a continuous
spin-field BP exchange model, as well as the effect of the discreteness
of the lattice are derived in Section \ref{Sec_BP}. Numerical results
on biskyrmions in a lattice model of a centrosymmetric film of finite
thickness with the nearest-neighbor exchange interaction, PMA, DDI,
and the external magnetic field are presented in Section \ref{Sec_Numerical}.
Our results and suggestions for experiments are discussed in Section
\ref{Sec_Discussion}.

\section{Skyrmions and biskyrmions in the 2D exchange model}

\label{Sec_BP}

\subsection{Spin field in a Belavin-Polyakov (BP) biskyrmion}

We begin with a 2D exchange model with the energy 
\begin{equation}
E_{ex}=\frac{J}{2}\int dxdy\left(\bm{\nabla}s_{\alpha}\cdot\bm{\nabla}s_{\alpha}\right),\label{Energy_exchange_continuous}
\end{equation}
where summation over spin components $\alpha=x,y,z$ is assumed. Here
$J$ is the exchange constant and ${\bf s}$ is a three-component
fixed-length spin field, ${\bf s}^{2}=1$. All spin-field configurations
${\bf s}(x,y)$ are divided into homotopy classes characterized by
the topological charge \cite{BelPolJETP75} 
\begin{equation}
Q=\int\frac{dxdy}{4\pi}\:{\bf s}\cdot\frac{\partial{\bf s}}{\partial x}\times\frac{\partial{\bf s}}{\partial y}\label{Q}
\end{equation}
that takes quantized values $Q=0,\pm1,\pm2,...$. The extremal spin-field
configurations satisfy 
\begin{equation}
\mathbf{s}\times\bm{\nabla}^{2}\mathbf{s}=0.\label{eq-motion}
\end{equation}
Throughout this paper we choose uniform magnetization, ${\bf s}=(0,0,-1)$,
in the negative z-direction at infinity.

The absolute energy minimum inside each homotopy class is given by
\begin{equation}
E_{ex}=4\pi J|Q|.\label{Energy_Q}
\end{equation}
The corresponding solutions with $Q>0$ are called skyrmions, while
solutions with $Q<0$ are called antiskyrmions. They have the simplest
form \cite{BelPolJETP75,Lectures} in terms of a complex variable
$\omega=\omega_{1}+i\omega_{2}$ with 
\begin{equation}
\omega_{1}=\frac{2s_{x}}{1+s_{z}},\qquad\omega_{2}=\frac{2s_{y}}{1+s_{z}}.\label{omega_def}
\end{equation}
Eq.\ (\ref{eq-motion}) then reduces to $\partial\omega/\partial x=\mp i\partial\omega/\partial y$
or, equivalently, to 
\begin{equation}
\frac{\partial\omega_{1}}{\partial x}=\pm\frac{\partial\omega_{2}}{\partial y},\qquad\frac{\partial\omega_{1}}{\partial y}=\mp\frac{\partial\omega_{2}}{\partial x}\label{omega-equations}
\end{equation}
that corresponds to the Cauchy-Riemann conditions for the complex
function $\omega$. They are satisfied by any analytical function
$\omega(z)$, where $z=x+iy$.

The minimum-energy solutions for a skyrmion with $Q=1$ and an antiskyrmion
with $Q=-1$ are given by $\omega=z/l$ and $\omega=z^{*}/l$, respectively,
with $z^{*}$ being the complex conjugate of $z$. The minimum-energy
biskyrmion with $Q=2$ corresponds to 
\begin{equation}
\omega(z)=e^{i\gamma}\frac{(z-d/2)(z+d/2)}{l^{2}}=e^{i\gamma}\frac{[z^{2}-(d/2)^{2}]}{l^{2}}.\label{biskyrmion_via_omega}
\end{equation}
It is parametrized by the chirality angle $\gamma$ and two lengths:
the parameter $l$ that roughly describes the size of the biskyrmion,
and another parameter $d$ that can be visually interpreted as the
separation of two $Q=1$ skyrmions in a biskyrmion, see Fig.\ \ref{Fig-biskyrmion}.
Notice that the nonlinearity of Eq.\ (\ref{eq-motion}) does not
support this interpretation per se, rather Eq.\ (\ref{biskyrmion_via_omega})
suggests that a biskyrmion is a product of two $Q=1$ skyrmions.

In terms of $\omega$ of Eq. (\ref{omega_def}) the components of
${\bf s}$ are given by 
\begin{equation}
s_{x}=\frac{{\rm Re}(\omega)}{1+|\omega|^{2}/4},\quad s_{y}=\frac{{\rm Im}(\omega)}{1+|\omega|^{2}/4},\quad s_{z}=\frac{1-|\omega|^{2}/4}{1+|\omega|^{2}/4}.\label{s_via_omega}
\end{equation}
These formulas allow one to obtain the spatial dependence of the components
of the spin field in the BP biskyrmion: 
\begin{eqnarray}
s_{x} & = & \frac{2\lambda^{2}(x^{2}-y^{2}-d^{2}/4)\cos\gamma-4\lambda^{2}xy\sin\gamma}{\lambda^{4}+(x^{2}+y^{2}-d^{2}/4)^{2}+d^{2}y^{2}}\label{biskyrmion_sx}\\
s_{y} & = & \frac{2\lambda^{2}(x^{2}-y^{2}-d^{2}/4)\sin\gamma+4\lambda^{2}xy\cos\gamma}{\lambda^{4}+(x^{2}+y^{2}-d^{2}/4)^{2}+d^{2}y^{2}}\label{biskyrmion_sy}\\
s_{z} & = & \frac{\lambda^{4}-(x^{2}+y^{2}-d^{2}/4)^{2}-d^{2}y^{2}}{\lambda^{4}+(x^{2}+y^{2}-d^{2}/4)^{2}+d^{2}y^{2}}\label{biskyrmion_sz}
\end{eqnarray}
where we have replaced $l$ with $\lambda$ satisfying $\lambda^{4}=4l^{4}$
to have fewer numerical factors in the formulas.

Fig.\ \ref{Fig-biskyrmion} provides visualization of the spin field
given by the above equations with $\gamma=\pi/2$ (Bloch-type biskyrmion).
At the saddle point $x=y=0$ one has 
\begin{equation}
s_{z}=s_{z,\mathrm{sad}}=\frac{\lambda^{4}-(d/2)^{4}}{\lambda^{4}+(d/2)^{4}}.\label{szsad}
\end{equation}
For $d\gg\lambda$, the BP biskyrmion becomes a superposition of two
$Q=1$ BP skyrmions with the size $\tilde{\lambda}=\lambda^{2}/d\ll\lambda$.

For a biantiskyrmion Eq. (\ref{biskyrmion_via_omega}) with $z\rightarrow z^{*}$
yields the expression similar to Eqs. (\ref{biskyrmion_sx})-(\ref{biskyrmion_sz})
with the sign of the $xy$ terms changed.

At $d=0$ Eqs. (\ref{biskyrmion_sx})-(\ref{biskyrmion_sz}) acquire
a simple form in the polar coordinates, ${\bf r}=(r,\phi)$, 
\begin{equation}
{\bf s}({\bf r})=\left\{ \frac{2\lambda^{2}r^{2}\cos(\gamma\pm2\phi)}{\lambda^{4}+r^{4}},\frac{2\lambda^{2}r^{2}\sin(\gamma\pm2\phi)}{\lambda^{4}+r^{4}},\frac{\lambda^{4}-r^{4}}{\lambda^{4}+r^{4}}\right\} ,\label{Skyrmion_Q=00003D00003D00003D2}
\end{equation}
with a plus sign for the $Q=2$ skyrmion and a minus sign for the
$Q=-2$ antiskyrmion.

In a particular case of a compact single-centered $|Q|=2$ topological
defect, that does not split into $Q=\pm1$ skyrmions or antiskyrmions
and is given by 
\begin{equation}
\omega=e^{i\gamma}\left(z/l\right)^{|Q|},\qquad\omega=e^{i\gamma}\left(z^{*}/l\right)^{|Q|},\label{omega-Q}
\end{equation}
respectively, one arrives at a more general expression that is valid
for an arbitrary positive or negative integer $Q$: 
\begin{eqnarray}
{\bf s}({\bf r}) & = & \left\{ \frac{2\lambda^{|Q|}r^{|Q|}\cos(\gamma+Q\phi)}{\lambda^{2|Q|}+r^{2|Q|}},\frac{2\lambda^{|Q|}r^{|Q|}\sin(\gamma+Q\phi)}{\lambda^{2|Q|}+r^{2|Q|}},\right.\nonumber \\
 &  & \left.\frac{\lambda^{2|Q|}-r^{2|Q|}}{\lambda^{2|Q|}+r^{2|Q|}}\right\} .\label{Skyrmion_general_Q}
\end{eqnarray}
To determine the size $\lambda$ of such a skyrmion from the numerically
computed spin field, one can use the integrals that can be obtained
with the help of Eq. (\ref{Skyrmion_general_Q}): 
\begin{equation}
I_{1}=S_{z}=2\pi\intop rdr\left[s_{z}(r)+1\right]=\frac{2\pi^{2}\lambda^{2}}{|Q|\sin\left(\pi/|Q|\right)}\label{Sz_Q>2}
\end{equation}
for $|Q|\geq2$ and 
\begin{equation}
I_{2}=2\pi\intop rdr\left[s_{z}(r)+1\right]^{2}=\frac{4\pi^{2}\lambda^{2}\left(|Q|-1\right)}{Q^{2}\sin\left(\pi/|Q|\right)},
\end{equation}
etc. Note that $I_{1}$ is the total spin, $S_{z}$, of the topological
defect (see below). If the values of $\lambda_{n}$ obtained through
different $I_{n}$ are close to each other, the shape of the skyrmion
is close to the BP shape. For $Q=\pm1$ the first integral logarithmically
diverges and one has to use $I_{n}$ with higher powers of $n$ to
characterize the shape of the bubble (see, e.g., Eq. (8) of Ref. \onlinecite{AP-2018}).

The energy of any spin configuration can be computed with the help
of Eq.\ (\ref{Energy_exchange_continuous}) or by noticing its equivalent
form 
\begin{equation}
E_{ex}=\frac{J}{2}\int dxdy\frac{|d\omega/dz|^{2}}{(1+|\omega|^{2}/4)^{2}}.\label{Energy_via_omega}
\end{equation}
Substitution of Eqs.\ (\ref{biskyrmion_via_omega}) or (\ref{omega-Q})
into Eq.\ (\ref{Energy_via_omega}) and integration with $dxdy=rdrd\phi$
and $z=x+iy=re^{i\phi}$reproduces Eq.\ (\ref{Energy_Q}).

\subsection{Magnetic moment of the BP biskyrmion}

The magnetic moment of the topological defect equals $g\mu_{B}S_{z}$,
where $g$ is the gyromagnetic factor, $\mu_{B}$ is the Bohr magneton
and $S_{z}$ is the total spin of the defect defined as the difference
between the spin of the $2D$ plane with and without the defect. For
the boundary condition $s_{z}=-1$ at infinity one has 
\begin{equation}
S_{z}=\int\frac{dxdy}{a^{2}}(s_{z}+1),\label{Sz_def}
\end{equation}
where $a$ is the lattice constant and $s_{z}$ is given by the last
of Eqs.\ (\ref{s_via_omega}), which yields 
\begin{equation}
s_{z}+1=\frac{2}{1+|\omega|^{2}/4}.\label{sz+1_via_omega}
\end{equation}

For a biskyrmion given by Eq.\ (\ref{biskyrmion_via_omega}) with
$l^{2}=\lambda^{2}/2$, switching to polar coordinates, one has 
\begin{equation}
|\omega|^{2}/4=u^{2}-2pu\cos(2\phi)+p^{2},\quad u\equiv\frac{r^{2}}{\lambda^{2}},\quad p\equiv\frac{d^{2}}{4\lambda^{2}}
\end{equation}
Substituting this into Eqs.\ (\ref{sz+1_via_omega}) and (\ref{Sz_def})
one obtains 
\begin{equation}
S_{z}=\left(\frac{\lambda}{a}\right)^{2}f(p),\label{Sz-final}
\end{equation}
where 
\begin{equation}
f(p)=\int_{0}^{\infty}\int_{0}^{2\pi}\frac{dud\phi}{u^{2}-2pu\cos(2\phi)+p^{2}+1}.
\end{equation}
Integration over the angle yields 
\begin{equation}
f(p)=2\pi\int_{0}^{\infty}\frac{du}{\left[{(u^{2}-p^{2}+1)^{2}+4p^{2}}\right]^{1/2}}\label{fp_integral}
\end{equation}

\begin{figure}[ht]
\centering{}\includegraphics[width=9cm]{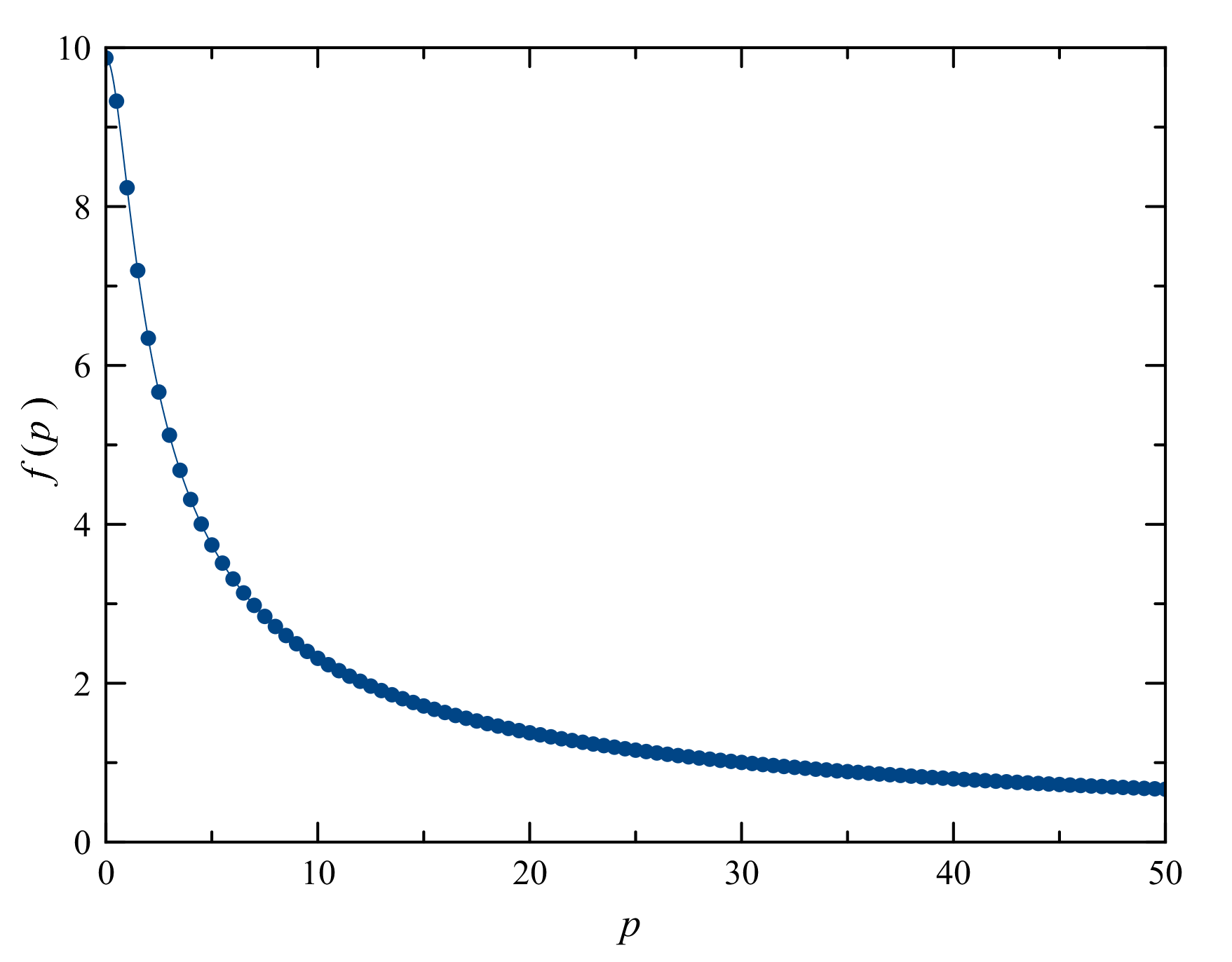} \caption{The plot of the function $f(p)$ given by Eq.\ (\ref{fp_integral}). }
\label{Fig-fp} 
\end{figure}

For an arbitrary $p$ this integral can be expressed via special functions
in a rather cumbersome way that we do not provide here. The function
$f(p)$ plotted numerically is shown in Fig.\ \ref{Fig-fp}. Its
asymptotic behavior can be computed analytically: 
\begin{equation}
S_{z}=\left(\frac{\lambda}{a}\right)^{2}\begin{cases}
\pi^{2}, & d=0\\
16\pi\left(\lambda/d\right)^{2}\ln(d/\lambda), & d\gg\lambda.
\end{cases}\label{Sz_biskyrmion_limits}
\end{equation}
Numerical evaluation gives a more accurate but close result, $\ln(1.1d/\lambda)$,
for the logarithmic cutoff in the case of $d\gg\lambda$.

According to Fig.\ \ref{Fig-fp} and Eqs.\ (\ref{Sz-final}), (\ref{Sz_biskyrmion_limits}),
at a fixed $\lambda$ the magnetic moment of the biskyrmion rapidly
decreases on increasing $d$. One should notice, however, that in
this limit the total spin of the biskyrmion can be written as 
\begin{equation}
S_{z}=8\pi\left(\frac{\tilde{\lambda}}{a}\right)^{2}\ln\left(\frac{1.05d}{\tilde{\lambda}}\right)
\end{equation}
in terms of the effective size $\tilde{\lambda}=\lambda^{2}/d\ll\lambda$
of the $Q=1$ skyrmion in a biskyrmion. Note that the magnetic moment
of an isolated $Q=1$ skyrmion is given by \cite{quantum} 
\begin{equation}
S_{z}^{\{1\}}=4\pi\left(\frac{\tilde{\lambda}}{a}\right)^{2}\ln\left(\frac{\Lambda}{\tilde{\lambda}}\right),
\end{equation}
where $\Lambda$ is a long-distance cutoff determined by the lateral
dimension of the film, $L$, or by $\delta_{H}=\sqrt{J/H}$, whichever
is shorter. Thus, by order of magnitude, $S_{z}$ remains the same
regardless of the separation of $Q=1$ skyrmions in the biskyrmion
and similar to the magnetic moment of an isolated $Q=1$ skyrmion.
As we shall see the energy minimum of a biskyrmion in a real magnetic
film is realized at $d\sim\lambda$. In this case $S_{z}\propto d^{2}\sim\lambda^{2}$,
that is, the magnetic moment is proportional to the area occupied
by the biskyrmion.

\subsection{Lattice-discreteness correction to the energy}

\label{subsec:Lattice_discreteness}

The scale invariance of the skyrmion energy is broken by the discreteness
of the lattice. The exchange energy of the atomic spins considered
as classical spin vectors with $\left|\mathbf{s}_{i}\right|=1$ has
the form 
\begin{equation}
{\cal H}_{ex}=-\frac{1}{2}\sum_{ij}J_{ij}{\bf s}_{i}\cdot{\bf s}_{j},\label{Ham_exchange}
\end{equation}
where $J_{ij}=J$ for the nearest neighbors and zero otherwise. We
use the simple cubic lattice. The corresponding correction to the
energy can be obtained by calculating this lattice sum using the BP
solution. For $\lambda\gtrsim a$, expanding the dot products to the
lowest second order in spatial derivatives of the spin field, one
obtains the well-known result 
\[
E_{ex}=\frac{J}{2}\int dxdy\left[\left(\partial_{x}\mathbf{s}\right)^{2}+\left(\partial_{y}\mathbf{s}\right)^{2}\right]
\]
that is equivalent to Eq.\ (\ref{Energy_exchange_continuous}) and
leads to Eq.\ (\ref{Energy_Q}). Expansion to the fourth order yields
the energy correction 
\begin{equation}
\delta E_{ex}=-\frac{Ja^{2}}{24}\int dxdy\left[\left(\partial_{x}^{2}\mathbf{s}\right)^{2}+\left(\partial_{y}^{2}\mathbf{s}\right)^{2}\right].\label{forthorder}
\end{equation}
For a singled-centered skyrmion with an arbitrary $Q$, substituting
Eq.\ (\ref{Skyrmion_general_Q}) into Eq.\ (\ref{forthorder}),
we obtain with the help of computer algebra 
\begin{equation}
\delta E_{ex}=-\frac{\pi^{2}J}{3}\frac{Q^{2}-1}{\sin(\pi/|Q|)}\left(\frac{a}{\lambda}\right)^{2}.
\end{equation}
The previously obtained result in Ref.\ \onlinecite{CCG-PRB2012}
for $Q=\pm1$, 
\begin{equation}
\delta E_{ex}=-\frac{2\pi J}{3}\left(\frac{a}{\lambda}\right)^{2},
\end{equation}
follows from this formula in the limit of $|Q|\rightarrow1$.

\section{Biskyrmions in a magnetic film with perpendicular magnetic anisotropy}

\label{Sec_Numerical}

\subsection{Lattice model and dipolar field}

In the numerical work we study the lattice model of a ferromagnetic
film of finite thickness with the energy given by the sum over lattice
sites $i,j$ 
\begin{equation}
{\cal H}={\cal H}_{ex}-H\sum_{i}s_{iz}-\frac{D}{2}\sum_{i}s_{iz}^{2}-\frac{E_{D}}{2}\sum_{ij}\Phi_{ij,\alpha\beta}s_{i\alpha}s_{j\beta}.\label{Hamiltonian}
\end{equation}
Here ${\cal H}_{ex}$ is given by Eq. (\ref{Ham_exchange}), $D$
is the easy-axis PMA constant, and $H\equiv g\mu_{B}SB$, with $S$
being the value of the atomic spin and $B$ being the induction of
the applied magnetic field. In the DDI part of the energy 
\begin{equation}
\Phi_{ij,\alpha\beta}\equiv a^{3}r_{ij}^{-5}\left(3r_{ij,\alpha}r_{ij,\beta}-\delta_{\alpha\beta}r_{ij}^{2}\right),\label{DDI}
\end{equation}
where $\mathbf{r}_{ij}\equiv\mathbf{r}_{i}-\mathbf{r}_{j}$ is the
displacement vector between the lattice sites and $\alpha,\beta=x,y,z$
denote cartesian components. The parameter $E_{D}=\mu_{0}M_{0}^{2}a^{3}/(4\pi)$
defines the strength of the DDI, with $M_{0}=g\mu_{B}S/a^{3}$ being
the magnetization for our lattice model and $\mu_{0}$ being the magnetic
permeability of vacuum.

The ratio of the PMI and DDI is given by the dimensionless parameter
$\beta\equiv D/(4\pi E_{D})$. For $\beta>1$, the energy of the uniform
state with spins directed along the $z$-axis is lower than that of
the state with spins lying in the film's plane. For $\beta<1$, the
state with spins in the plane has a lower energy. The most interesting
practical case is $\beta\sim1$ realized in many materials in which
there is a considerable compensation of the effects of the PMA and
DDI.

In most materials the exchange interaction is much stronger than all
other interactions of the spins. Consequently in a centrosymmetric
system with magnetic anisotropy and nonsingular defects, such as skyrmions,
the magnetization can only smoothly rotate on a large spatial scale
gauged by two lengths: the domain-wall width, $\delta=a\sqrt{J/D}\gg a$,
and another length, $\delta_{H}=a\sqrt{J/H}\gg a$, generated by the
field. To describe such non-uniform spin states one needs to consider
a system with macroscopically large number of spins. In the lattice
model this can lead to impractically large computation times. Besides,
the systems with interactions that are weak compared to the ferromagnetic
exchange are magnetically soft because the energy of the non-uniform
structures per spin is small. This makes the convergence of the energy-minimization
routine very slow. The good news is that in such a case the atomic-scale
spatial resolution is excessive.

To speed up the computation for structures that are much larger than
the atomic spacing $a$ one can rescale the problem to another lattice
constant $b>a$ by first rewriting the energy in the continuous approximation
and then discretizing it again. The rescaled model with the parameters
\begin{equation}
J'=\frac{b}{a}J,\quad H'=\frac{b^{3}}{a^{3}}H,\quad D'=\frac{b^{3}}{a^{3}}D,\quad E'_{D}=\frac{b^{3}}{a^{3}}E_{D}\label{rescaling}
\end{equation}
has smaller number of mesh points $N_{\alpha}'=(a/b)N_{\alpha}$ and
smaller mismatch between $J$ and other parameters. This provides
faster convergence. After the computation with the rescaled model
is completed, one obtains the results for the original system by rescaling
the parameters back. Note that the domain-wall width and the field
related length are the same in the original and rescaled models, 
\begin{equation}
\delta'=b\sqrt{\frac{J'}{D'}}=a\sqrt{\frac{J}{D}}=\delta,\quad\delta'_{H}=b\sqrt{\frac{J'}{H'}}=a\sqrt{\frac{J}{H}}=\delta_{H}.
\end{equation}
The same is valid for all spatial structures such as magnetic bubbles,
etc. The above-mentioned rescaling is only important for the study
of films of large lateral dimensions. In a confined geometry, such
as, e.g., nanotracks, one can perform computation at the atomic level.

The use of strong anisotropy and strong DDI allows one to work with
a relatively small system and to have a reasonably fast convergence
\cite{GCZ-EPL2017}. In this case, however, the shape of the topological
defects is closer to that of the thin-wall skyrmion bubbles than to
the BP skyrmions. To obtain the latter one has to work with the skyrmion
size $\lambda$ satisfying $a\lesssim\lambda\lesssim\delta$ that
requires a smaller anisotropy constant. To investigate biskyrmions,
which is the purpose of this paper, we use $D/J=0.001$. This requires
a large system size, $N_{x}\times N_{y}=1000\times1000$, and results
in longer computation times. 
\begin{figure}
\begin{centering}
\includegraphics[width=9cm]{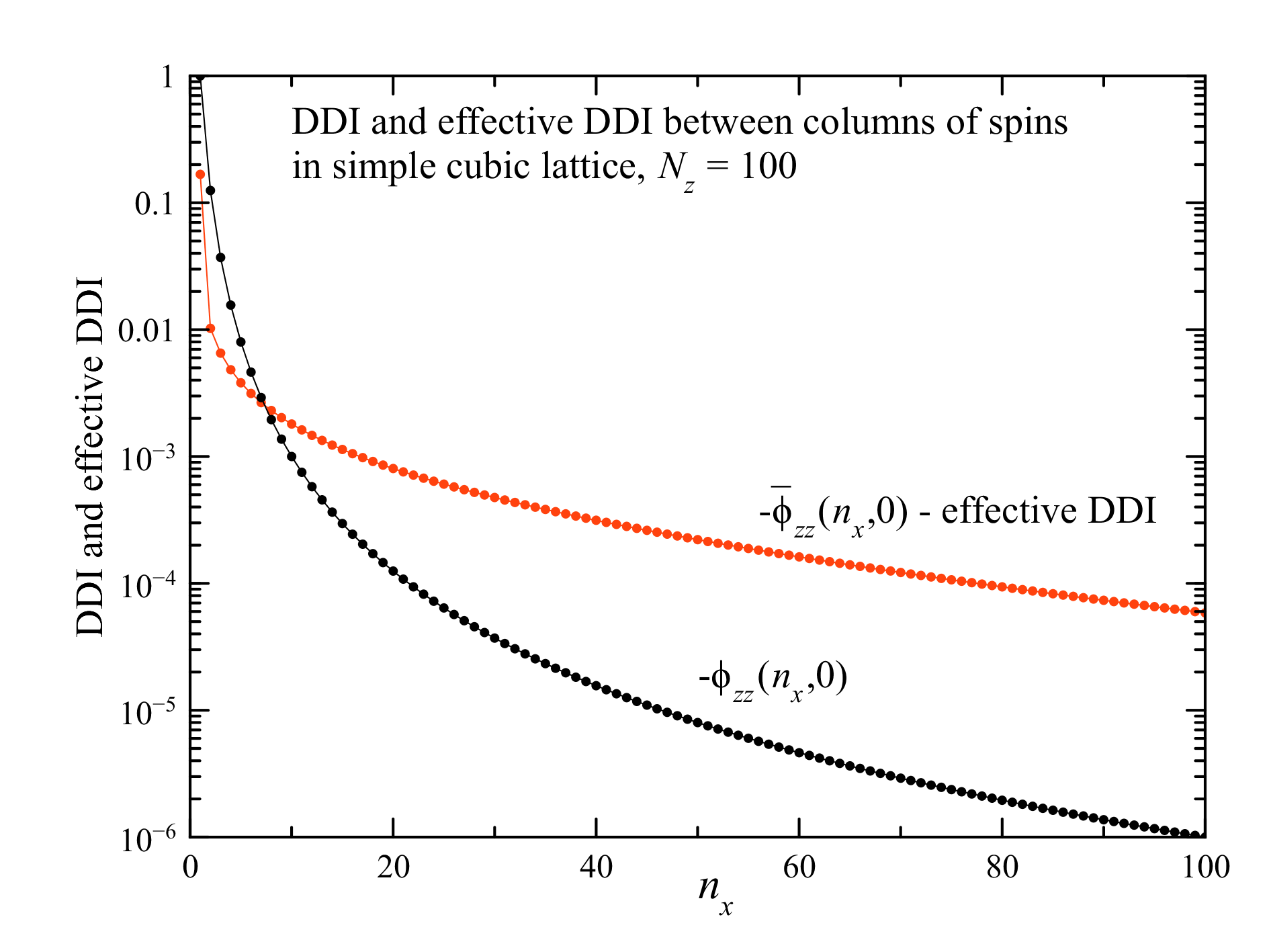} 
\par\end{centering}
\begin{centering}
\includegraphics[width=9cm]{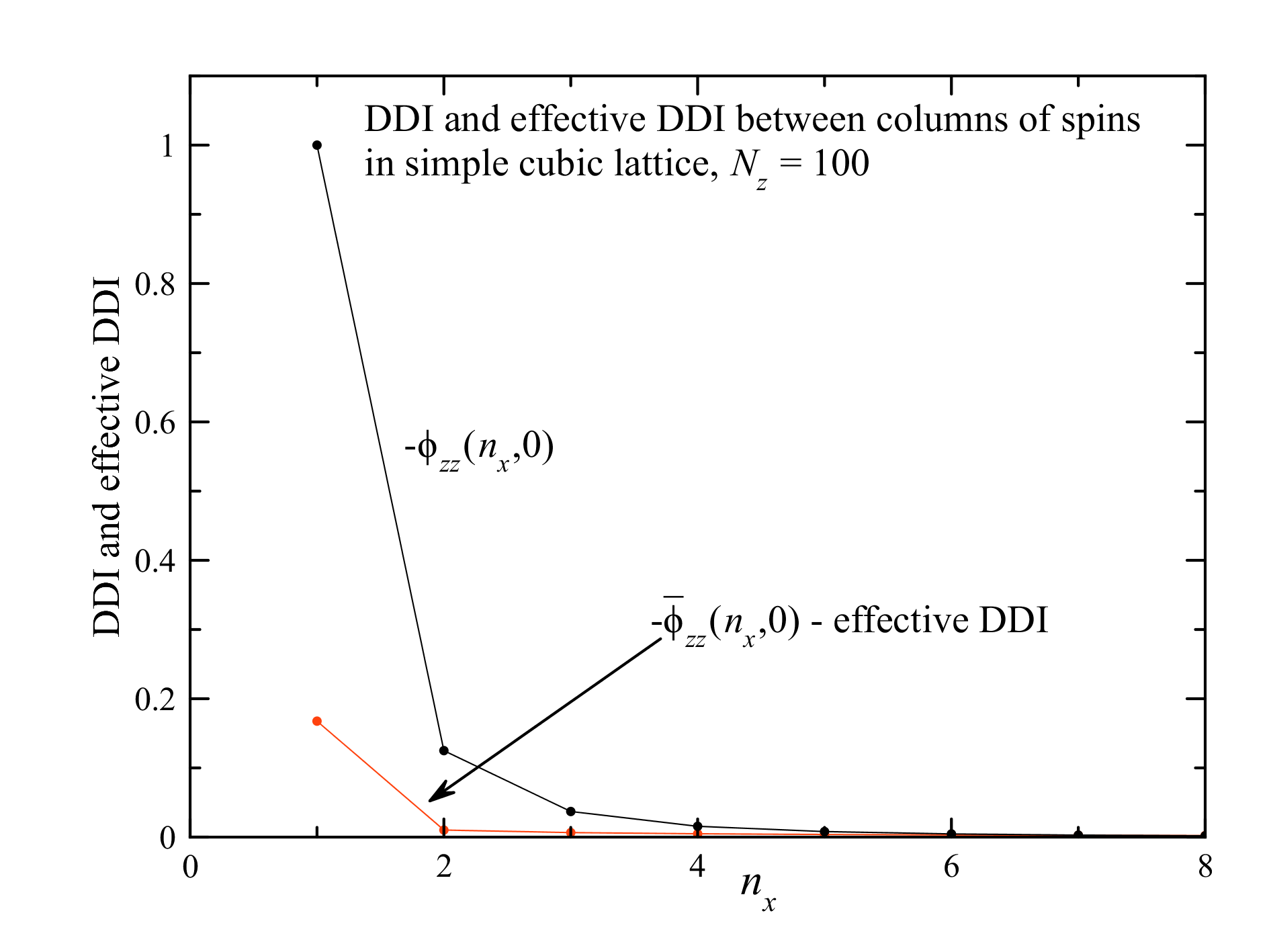} 
\par\end{centering}
\caption{DDI and effective DDI between columns of spins in a simple cubic lattice
for the film thickness $N_{z}=100$. Upper panel: Logarithmic scale.
Lower Panel: Linear scale at short distances.}
\label{Fig-Vzz_eff_vs_dist} 
\end{figure}

An important parameter controlling the DDI is the film thickness represented
by $N_{z}$ in the units of the atomic spacing $a$. For thin films
that are studied here, the magnetization inside the film is nearly
constant along the direction perpendicular to the film. Thus one can
make the problem effectively two-dimensional by introducing the effective
DDI between the columns of parallel spins, considered as effective
spins of the 2D model. This greatly speeds up the computation. To
this end, for the simple cubic lattice, one can write the dipolar
coupling, Eq. (\ref{DDI}), as $\varPhi_{ij,\alpha\beta}=\phi_{\alpha\beta}\left(n_{x},n_{y},n_{z}\right)$,
where $n_{x}\equiv i_{x}-j_{x}$ etc., are the distances on the lattice
and 
\begin{equation}
\phi_{\alpha\beta}\left(n_{x},n_{y},n_{z}\right)=\frac{3n_{\alpha}n_{\beta}-\delta_{\alpha\beta}\left(n_{x}^{2}+n_{y}^{2}+n_{z}^{2}\right)}{\left(n_{x}^{2}+n_{y}^{2}+n_{z}^{2}\right)^{5/2}}.\label{phinnnDef-1}
\end{equation}
The effective DDI is defined by 
\begin{equation}
\bar{\phi}_{\alpha\beta}\left(n_{x},n_{y}\right)=\frac{1}{N_{z}}\sum_{i_{z},j_{z}=1}^{N_{z}}\phi_{\alpha\beta}\left(n_{x},n_{y},i_{z}-j_{z}\right).
\end{equation}
Using the symmetry, one can express this result in the form with only
one summation, 
\begin{eqnarray}
\bar{\phi}_{\alpha\beta}\left(n_{x},n_{y}\right) & = & \phi_{\alpha\beta}\left(n_{x},n_{y},0\right)\\
 & + & \frac{2}{N_{z}}\sum_{n_{z}=1}^{N_{z}-1}\left(N_{z}-n_{z}\right)\phi_{\alpha\beta}\left(n_{x},n_{y},n_{z}\right),\nonumber 
\end{eqnarray}
that is used in the computations.

The effective DDI (that can be pre-computed) has different forms in
different ranges of the distance $r$. At $r\apprge aN_{z}$ it scales
as the interaction of magnetic dipoles $1/r^{3}$, while at $r\lesssim aN_{z}$
it goes as $1/r$ that corresponds to the interaction of magnetic
charges at the surface of the film. Numerical results in Fig. \ref{Fig-Vzz_eff_vs_dist}
show that at large distances the effective DDI is stronger in films
of finite thickness than in pure 2D systems. This has an important
effect on the stability of biskyrmions.

In the computations, the dipolar field from the downward spins outside
the system was added to obtain the results that are valid for an infinite
system. This field was computed as the field of the plate of a very
large size magnetized downward plus the dipolar field created by the
finite system under the consideration magnetized upward.

\subsection{Energy landscape of rigid-shape biskyrmions}

Interactions other than exchange deform BP skyrmions and biskyrmions.
As the result, the exchange energy increases. In the numerical work
this increase provides the measure of the shape distortion. Also,
one can consider the energy $\Delta E$ of the state with skyrmions
with respect to that of the uniform state with all spins down. This
energy is smaller than $4\pi J|Q|$ due to the contributions of the
PMA and DDI.

\begin{figure}
\begin{centering}
\includegraphics[width=9cm]{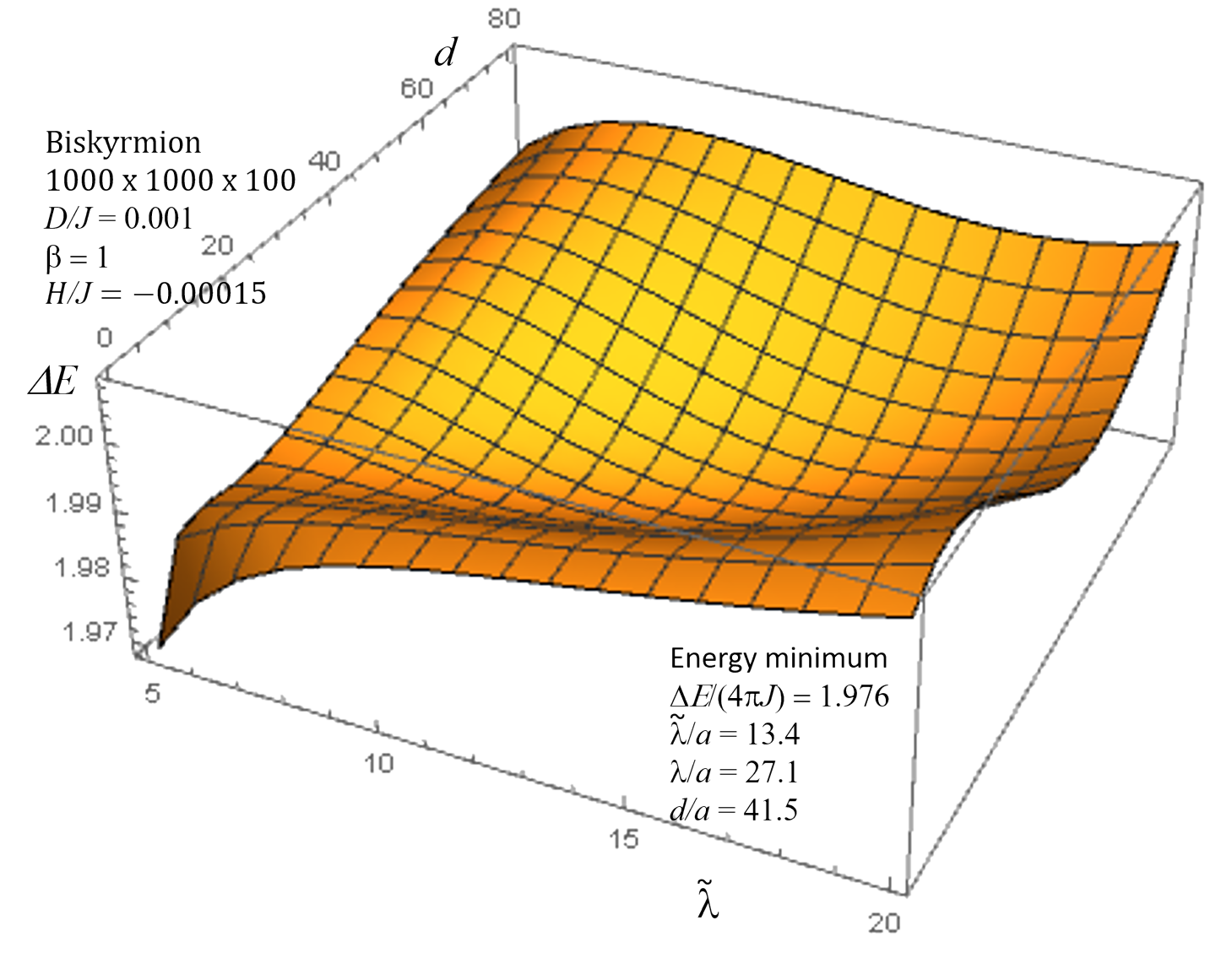} 
\par\end{centering}
\begin{centering}
\includegraphics[width=9cm]{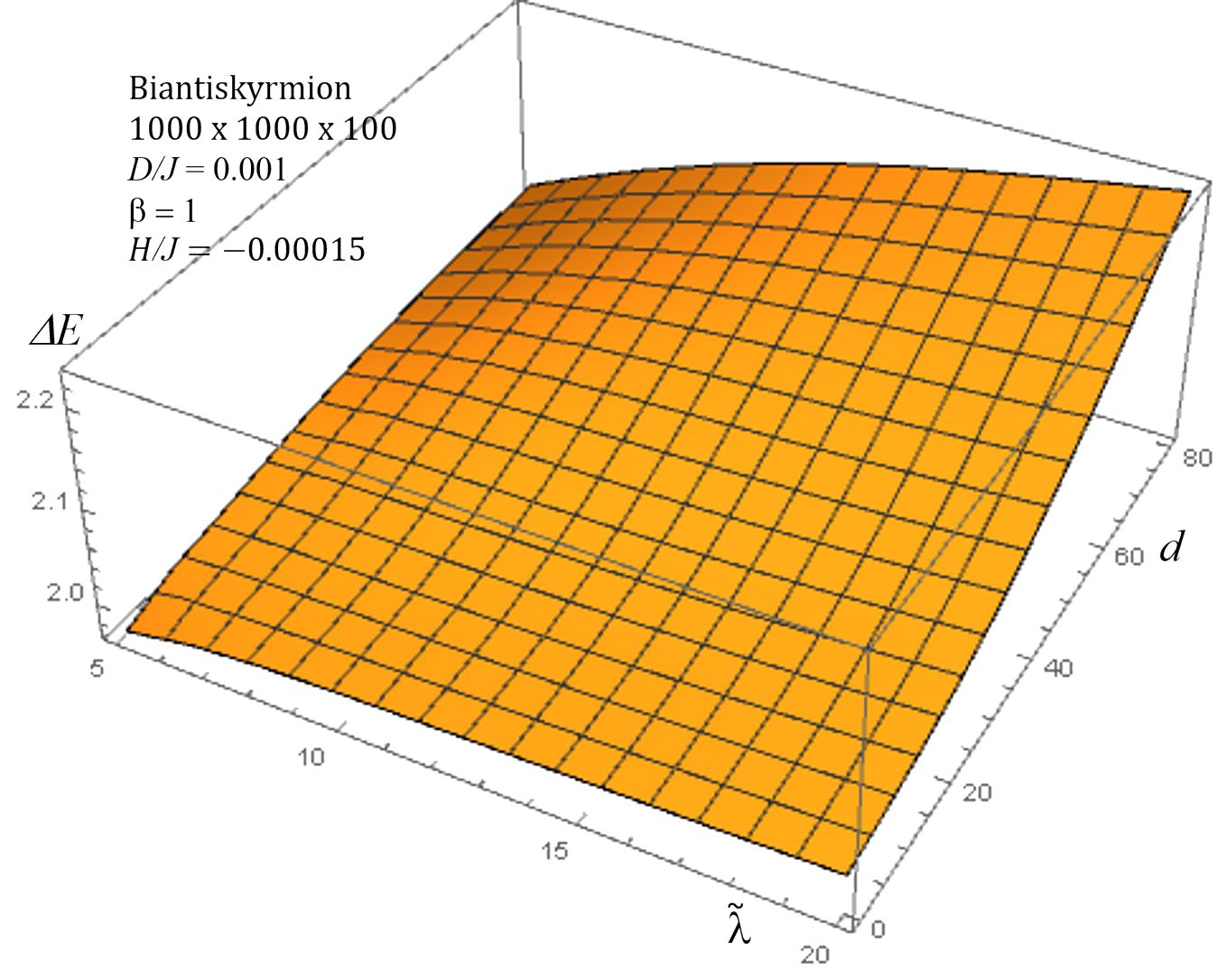} 
\par\end{centering}
\caption{The energy landscape of a Belavin-Polyakov biskyrmion (upper panel)
and biantiskyrmion (lower panel). The biskyrmion has a local energy
minimum at a finite separation, $d>0$, while the biantiskyrmion for
the same parameters has no energy minimum at all.}
\label{Fig-Energy_landscape} 
\end{figure}

We start with exploring the energy landscape, $E(\lambda,d$), in
the presence of biskyrmions, assuming the rigid BP shape given by
Eqs.\ (\ref{biskyrmion_sx})-(\ref{biskyrmion_sz}) and numerically
calculating the energy with the help of Eq.\ (\ref{Hamiltonian}).
This is valid when the exchange is much greater than all other interactions
and when the biskyrmion size is small compared to the domain wall
width: $\lambda,d\lesssim\delta$. In fact, it is more convenient
to parametrize $\lambda$ as $\lambda^{2}=\tilde{\lambda}(\tilde{\lambda}+d)$,
where $\tilde{\lambda}$ is the actual size of an individual $Q=1$
skyrmion in the limit of a large separation $d$.

Whereas in the pure exchange model the energy of biskyrmions is independent
of $\lambda$ and $d$, other interactions (here PMA, DDI, and Zeeman),
with the applied field $H$ as a control parameter, break this invariance
of the skyrmion energy and select the values of $\lambda$ and $d$
that provide the energy minimum. The energy minimum of the topological
defect with $Q=2$ always corresponds to a biskyrmion with $d\neq0$.

To the contrary, we find that for a bi-antiskymion the energy always
has a minimum at $d=0$, thus, there should be only $d=0$ single-centered
antiskyrmions with $Q=-2$. This confirms the findings of Ref. \onlinecite{GCZ-EPL2017},
see its Fig.\ 9 that shows different kinds of magnetic bubbles obtained
by the relaxation from a random spin state: All objects with $Q=-2$
are spatially symmetric antiskyrmions, whereas all objects with $Q=2$
have a finite separation, $d\neq0$.

\begin{figure}[ht]
\begin{centering}
\includegraphics[width=8cm]{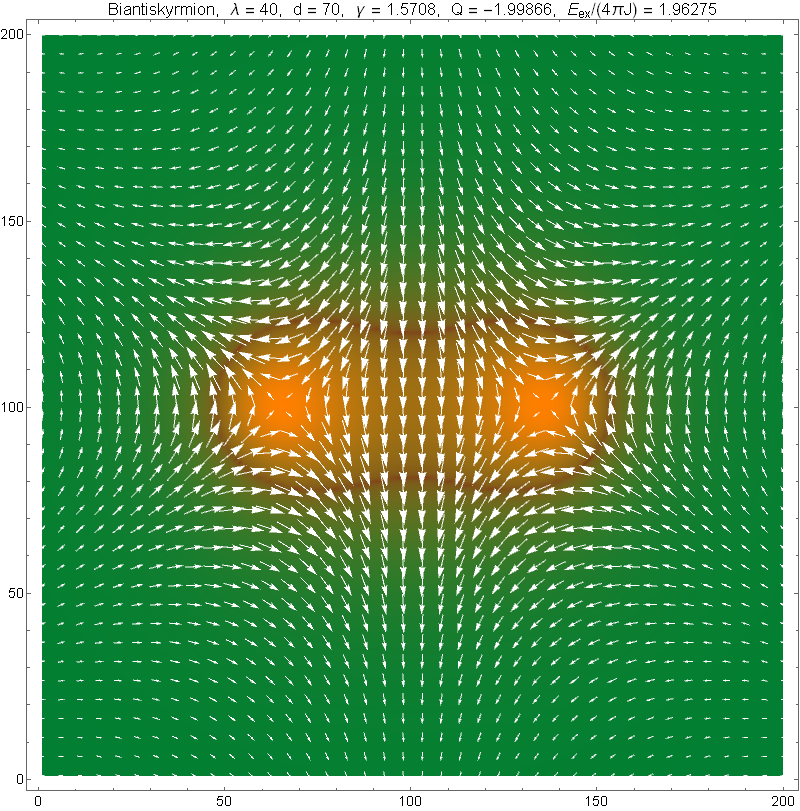} 
\par\end{centering}
\begin{centering}
\includegraphics[width=8cm]{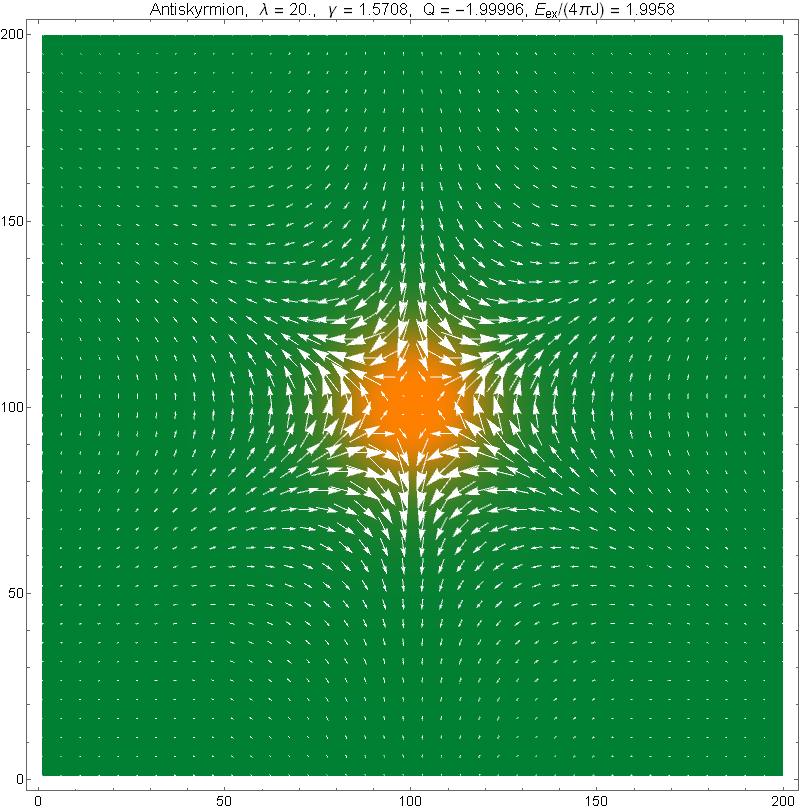}
\par\end{centering}
\centering{}\caption{Upper panel: Spin field in a $Q=-2$ BP biantiskyrmion ($d>0$) .
Lower panel: Spin field in a $Q=-2$ BP antiskyrmion ($d=0$) . In
the presence of DDI, PMA, and the applied magnetic field, the initial
state with $d>0$ in the course of relaxation evolves to the minimal-energy
state with $d=0$.}
\label{Fig-biantiskyrmion} 
\end{figure}

Fig.\ \ref{Fig-Energy_landscape} (upper panel) shows an example
of the biskyrmion energy landscape for the system of size $N_{x}\times N_{y}\times N_{z}=1000\times1000\times100$
for which most of the computations have been done. We used $D/J=0.001$
and $\beta=1$. For $H/J=-0.00015$ there is a local energy minimum
at $\lambda/a=27.1$ and $d/a=41.5$. The difference $\Delta E$ from
the energy of the uniformly magnetized state with spins down is given
in the units of $4\pi J$. For a weakly distorted BP biskyrmion it
is close to 2. As the parameter $\lambda$ decreases, the energy goes
down due the lattice-discreteness correction, see Sec. \ref{subsec:Lattice_discreteness}.
On the other hand, for the biantiskyrmion shown in Fig. \ref{Fig-biantiskyrmion}
(upper panel), for the same parameters as above, there is no energy
minimum at all, see lower panel in Fig.\ \ref{Fig-Energy_landscape}.
Thus the biantiskyrmion evolves to a $Q=-2$ antiskyrmion shown in
Fig. \ref{Fig-biantiskyrmion} (lower panel) It should be noted that
the computed energy landscape is the same for $\phi=0$ and $\phi=\pi/2$.
For smaller $|H|$, there is an energy minimum of the biantiskyrmion
at $d=0$ and $\lambda>0$. 

Whereas the rigid biskyrmion approximation provides a good qualitative
description, it is not completely accurate since the balance of different
interactions is very subtle and small deformations of the shape have
a significant effect on the energy. More accurate results can be obtained
by the numerical energy minimization described in the next section.

\subsection{Numerical energy minimization and results}

In this section we compute minimum-energy configurations of spins.
The numerical method \cite{DCP-PRB2013} combines sequential rotations
of spins ${\bf s}_{i}$ towards the direction of the local effective
field, ${\bf H}_{{\rm eff},i}=-\partial{\cal H}/\partial{\bf s}_{i}$,
with the probability $\alpha$, and the energy-conserving spin flips
(so-called \textit{overrelaxation}), ${\bf s}_{i}\to2({\bf s}_{i}\cdot{\bf H}_{{\rm eff},i}){\bf H}_{{\rm eff},i}/H_{{\rm eff},i}^{2}-{\bf s}_{i}$,
with the probability $1-\alpha$. The parameter $\alpha$ plays the
role of the effective relaxation constant. We mainly use the value
$\alpha=0.03$ that provides the overall fastest convergence.

The dipolar part of the effective field takes the longest time to
compute. The method uses Fast Fourier Transform (FFT) in the whole
sample as one program step. Since the dipolar field is much weaker
than the exchange, several cycles of spin alignment can be performed
before the dipolar field is updated, which increases the computation
speed. The total charge $Q$ of the topological defect has been computed
numerically using the lattice-discretized version of Eq.\ (\ref{Q}).

Computations were performed with Wolfram Mathematica using compilation.
Most of the numerical work has been done on the 20-core Dell Precision
T7610 Workstation. The FFT for computing the DDI was performed via
Mathematica's function ListConvolve that implicitly uses many processor
cores. For this reason, no explicit parallelization was done in our
program. However, we have been able to run several independent computations
at the same time.

\begin{figure}
\begin{centering}
\includegraphics[width=9cm]{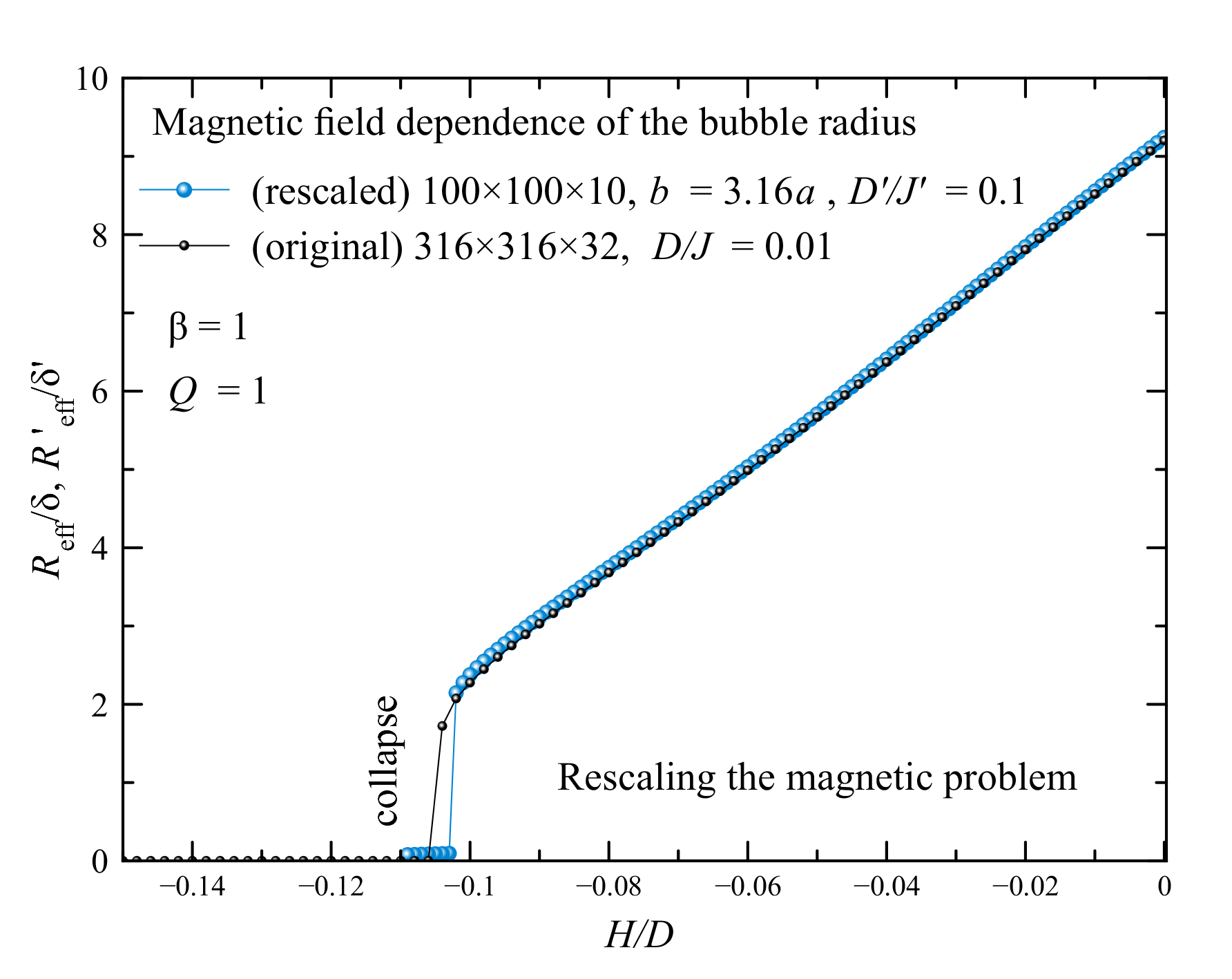} 
\par\end{centering}
\caption{Magnetic field dependence of the effective radus of a stable skyrmion
bubble for the ``original'' and rescaled models.}
\label{Fig-Reff} 
\end{figure}

To demonstrate how well our rescaling method given by Eq. (\ref{rescaling})
works, we have computed the effective radius $R_{\mathrm{eff}}$ of
topological defects with $Q=1$, defined as $\pi R_{\mathrm{eff}}^{2}=S_{z}/2$,
where $S_{z}$ is the total spin of the bubble defined by $S_{z}=\intop dxdy\left[s_{z}(x,y)+1\right]$.
The ``original'' system is a grid of $316\times316\times32$ spins
with $D/J=0.01$ and the lattice constant $a$. The rescaled system
is a grid of $100\times100\times10$ spins with $D'/J'=0.1$ and the
lattice constant $b=\sqrt{10}a=3.16a$. In both cases $\beta=1$.
Fig. \ref{Fig-Reff} shows a perfect agreement between $R_{\mathrm{eff}}$
and $R'_{\mathrm{eff}}$. (Note that $\delta'=\delta$ thus practically
$R'_{\mathrm{eff}}=R_{\mathrm{eff}}$) in the plot.

\begin{figure}
\begin{centering}
\includegraphics[width=9cm]{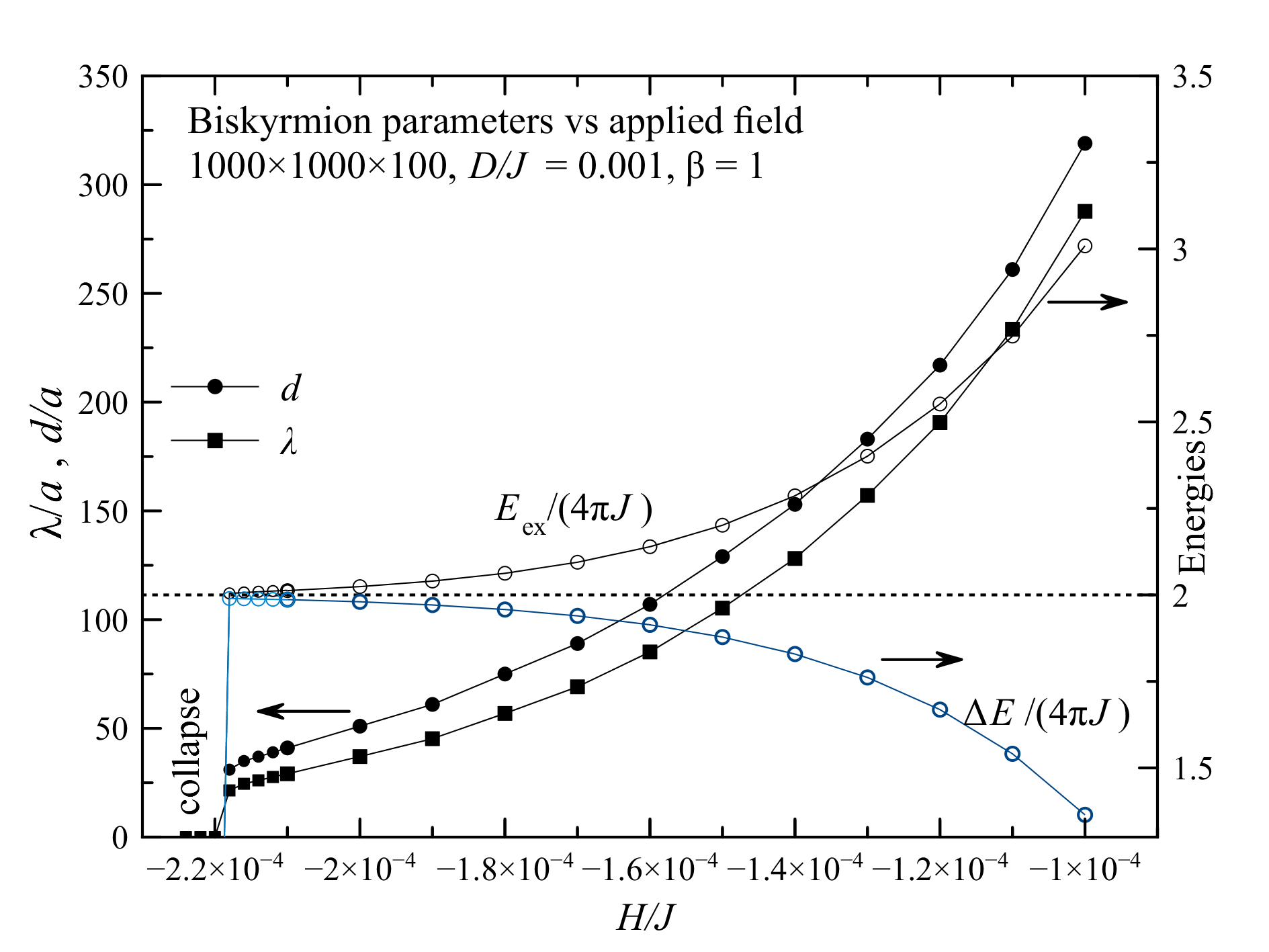} 
\par\end{centering}
\caption{Magnetic field dependences of the parameters $\lambda$ and $d$ of
stable biskyrmions in a film of thickness $N_{z}=100$ with $D/J=0.001$
and $\beta=1$. Exchange energy $E_{ex}$ and the total energy $\Delta E$
(with respect to the uniformly magnetized spins with spins down) are
shown on the right $y$-axis. For stronger fields, the biskyrmion
is smaller and close to the BP biskyrmion, as shown by the energies. }

\label{Fig-lam_d_vs_H-1} 
\end{figure}

Subsequently we performed the energy minimization for biskyrmions
with $Q=2$ in a system with $N_{x}\times N_{y}\times N_{z}=1000\times1000\times100$,
$D/J=0.001$, and $\beta=1$ at different values of the applied field
$H$. The separation $d$ was found numerically as the distance between
the maxima of $s_{z}$ in the biskyrmion. Then $\lambda$ was extracted
with the help of Eq.\ (\ref{szsad}) using the numerically found
saddle-point value of $s_{z}$.

\begin{figure}[ht]
\begin{centering}
\includegraphics[width=8cm]{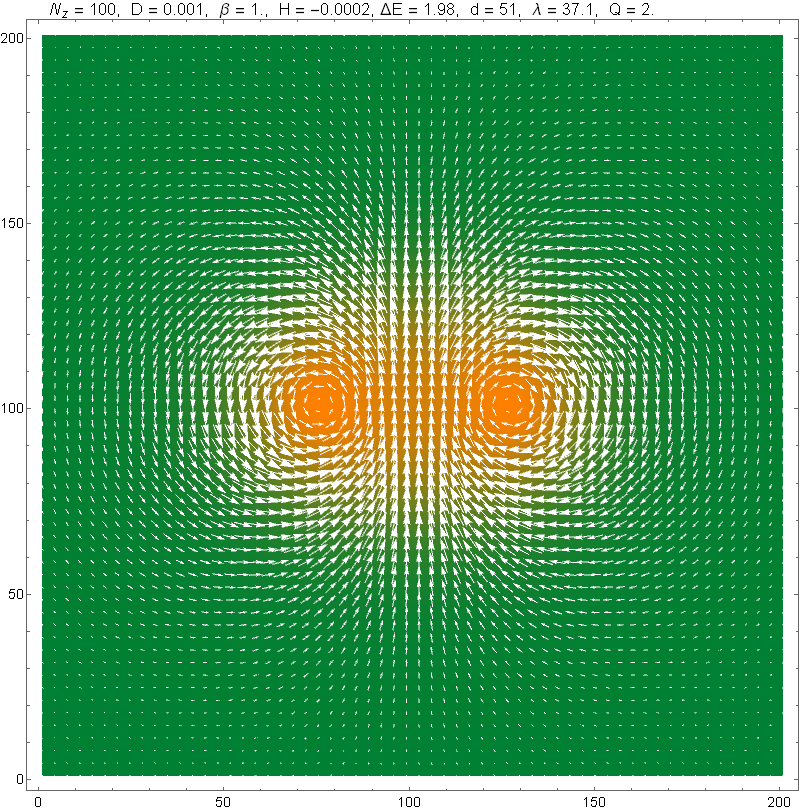} 
\par\end{centering}
\centering{}\includegraphics[width=8cm]{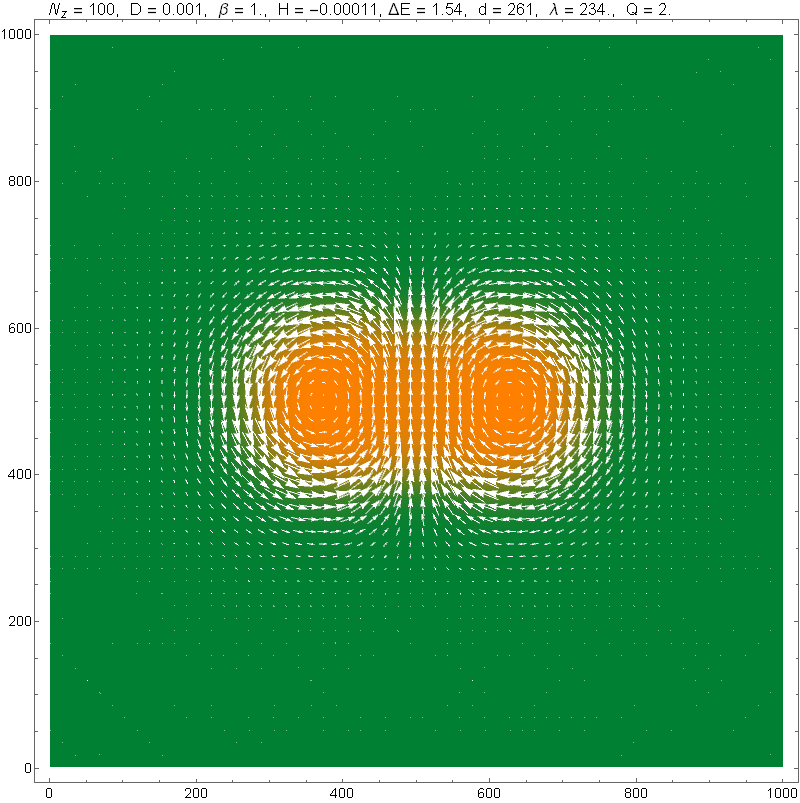} \caption{Computer-generated images of the spin field in a stable BP biskyrmion
(upper panel) and a biskyrmion bubble (lower panel) numerical solutions
of the centrosymmetric Heisenberg lattice model with the ferromagnetic
exchange, DDI, PMA, and external field. The values of parameters are
written above the figures. Orange/green color indicates positive/negative
$z$-component of the spin field. The in-plane spin components are
shown as white arrows.}
\label{Biskyrmion-Bubble} 
\end{figure}

In Fig.\ \ref{Fig-lam_d_vs_H-1} one can see that there is always
a finite separation, $d>0$, in the biskyrmion. For stronger fields,
the biskyrmion is smaller and closer to the BP shape, as the energies
$E_{ex}$ and $\Delta E$ are close to $8\pi J$. Biskyrmions collapse
as the field reaches the stability threshold. With decreasing the
magnitude of field the biskyrmion is expanding, gradually transforming
into a thin-wall bubble. The ratio $d/\lambda$ changes from 1.5 on
the left side of Fig.\ \ref{Fig-lam_d_vs_H-1} to 1.1 on its right
side. The transition from BP biskyrmions to biskyrmion bubbles can
be seen in a significant deviation of energy from $E/(4\pi J)=2$.
As the field strength further decreases, the bubble loses its circular
shape and transforms into a laminar domain.

Typical spin configurations of biskyrmions are illustrated (zoomed)
in Fig.\ \ref{Biskyrmion-Bubble} that provides more details as compared
to Fig.\ \ref{Fig-biskyrmion}. The in-plane spin components in a
BP biskyrmion shown in the upper panel of Fig.\ \ref{Biskyrmion-Bubble}
decay as $1/r^{2}$ in accordance with Eq.\ (\ref{biskyrmion_sx})
and Eq.\ (\ref{Skyrmion_Q=00003D00003D00003D2}). On the contrary,
in a biskyrmion bubble shown in the lower panel of Fig.\ \ref{Biskyrmion-Bubble}
the in-plane spin components decay exponentially and are hardly visible
away from the bubble. For topological defects with $Q=-2$ (antiskyrmions)
no finite separation $d$ was detected in our computations.

\begin{figure}
\begin{centering}
\includegraphics[width=9cm]{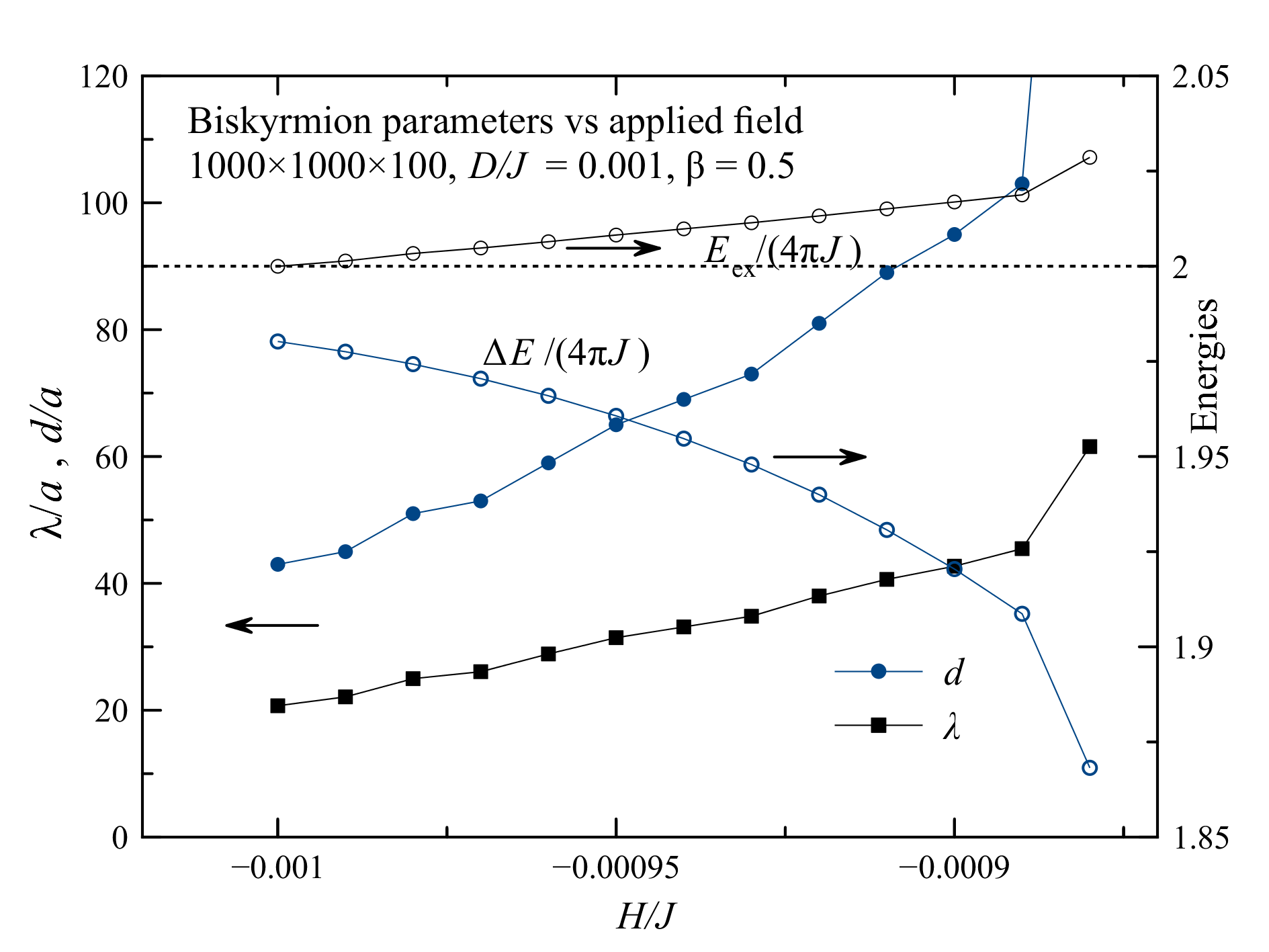} 
\par\end{centering}
\caption{Magnetic field dependences of the parameters $\lambda$ and $d$ of
stable biskyrmions in a film of thickness $N_{z}=100$ with $D/J=0.001$
and $\beta=0.5$.}

\label{Fig-lam_d_vs_H-2} 
\end{figure}

Next we performed computations for a similar model with a stronger
DDI, $\beta=0.5$. The DDI in excess of the PMA forces the spins into
the film's plane, which suppresses skyrmions. To prevent this from
happening, a stronger negative magnetic field has to be applied. The
results in Fig.\ \ref{Fig-lam_d_vs_H-2} show a larger separation,
$d/\lambda\apprge2$, and the shape close to the BP shape as $E_{ex}/(4\pi J)$
is rather close to 2. On the right side of the figure, the instability
of the uniform state with spins down occurs on decreasing the field's
strength.

For the model in which PMA is stronger than DDI, $\beta>1$, no stable
skyrmions or biskyrmions were found as they were collapsing even at
$H=0$. The topological structures can be stabilized by the negative
magnetic field in the range $\beta^{*}<\beta<1$ that depends on the
film's thickness $N_{z}$. Numerical studies show that the range of
$\beta$ narrows down for thin films. In the latter, the effect of
the DDI is similar to that of the easy-plane PMA, so that one can
introduce the effective anisotropy that includes both PMA and DDI,
$\widetilde{D}=D(1-1/\beta)$. This effective anisotropy changes its
sign at $\beta=1$. This results in the extremely weak distortion
of the BP shape of biskyrmions and extremely small controlling fields
$H$. However, the model with a single effective anisotropy cannot
support stable topological defects, including skyrmions and biskyrmions,
at any $\beta\neq1$. Since in real materials $\beta$ cannot be tuned,
very thin non-chiral films with no DMI are not a good medium for the
skyrmions. To the contrary, for thicker films, competition of the
short-range PMA and long-range ($\sim1/r$) DDI creates a range of
$\beta$ in which skyrmions and biskyrmions can exist.

\begin{figure}
\begin{centering}
\includegraphics[width=9cm]{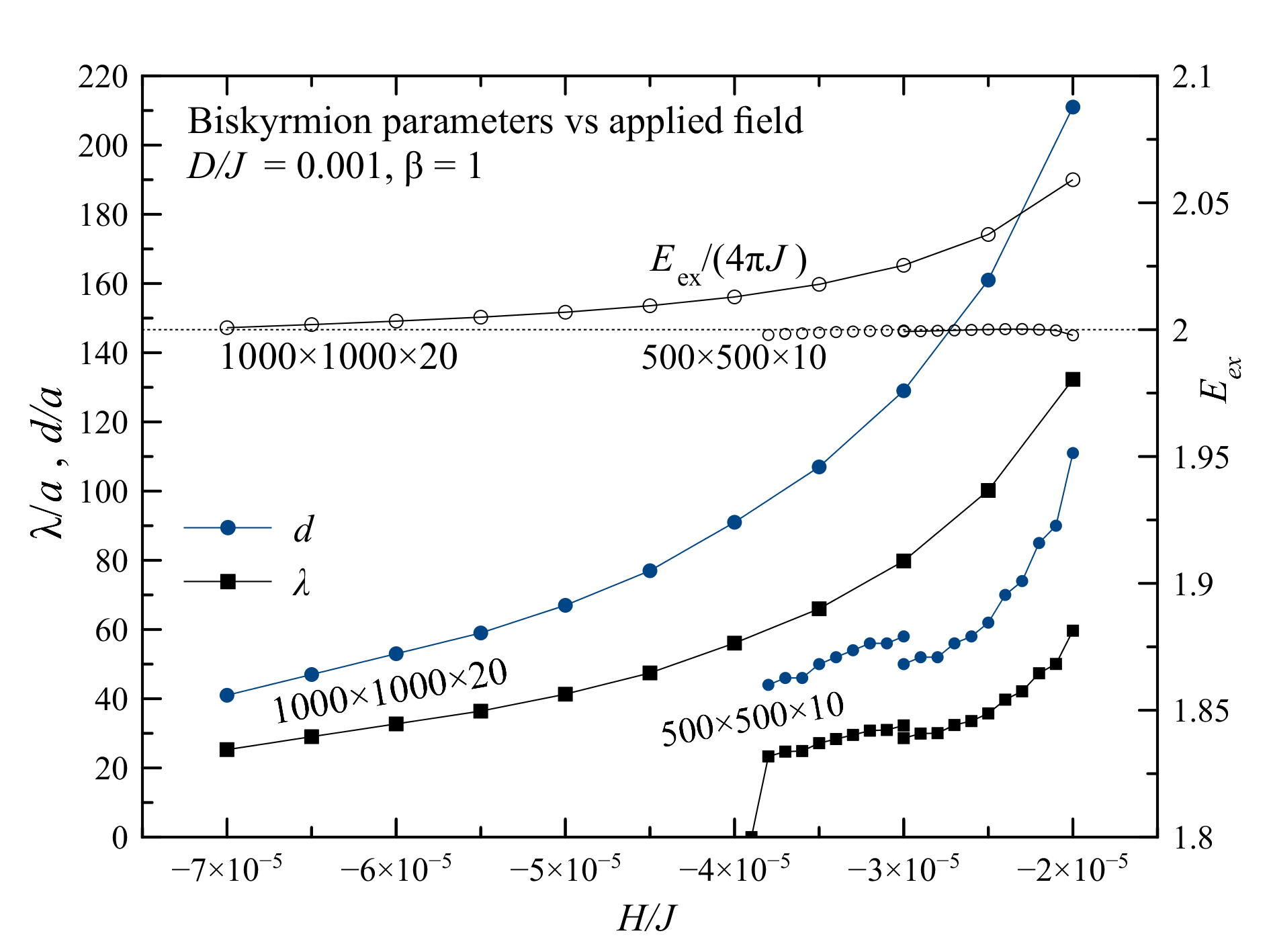} 
\par\end{centering}
\caption{Magnetic field dependences of the parameters $\lambda$ and $d$ of
stable biskyrmions (left axis) and the exchange energy (right axis)
in films of thickness $N_{z}=20$ and 10 with $D/J=0.001$ and $\beta=1.$}

\label{Fig-lam_d_vs_H-3} 
\end{figure}

In particular, for $N_{z}=20$ with $D/J=0.001$, stable biskyrmions
were found only for $\beta=1$. The results in Fig. \ref{Fig-lam_d_vs_H-3}
show very weakly distorted BP biskyrmions with practically the same
ratio $d/\lambda=1.62$ in the whole range of a very weak field $H$.
For $\beta=0.5$ and even for $\beta=0.75$ either the skyrmions collapse
if the negative field is too strong or the background spin-down state
becomes unstable if the negative field is too weak. For the system
of $500\times500\times10$ spins, biskyrmions were still found at
$\beta=1$ but the range of $H$ was much narrower. One can see that
here the exchange energy is much closer to 2 than for $N_{z}=20$,
thus the shape of the biskyrmion is much closer to the BP shape. For
$H/J<-3.8\times10^{-5}$ the biskyrmion collapses. For $H/J>-2\times10^{-5}$
the spin-down background becomes unstable.

Computations on the $500\times500$ monolayer found no biskyrmions
at all even for $\beta=1$. Initial states in the form of biskyrmions
evolve into states with separate skyrmions far away from each other
that exist in an extremely narrow region of $H$. 

The bottom line of this investigation is that biskyrmions should be
searched for in sufficiently thick films, such as $N_{z}=100$ for
$D/J=0.001$. For other values of the PMA, the scaling of Eq. (\ref{rescaling})
can be used. For a stronger PMA, biskyrmions can be supported by thinner
films.

\section{Discussion}

\label{Sec_Discussion}

We have studied biskyrmions in non-chiral ferromagnetic films of finite
thickness with perpendicular magnetic anisotropy (PMA) with account
of dipole-dipole interaction (DDI), and discreteness of the atomic
lattice. In agreement with experimental findings \cite{Yu-2014,Zhang2016}
we have found that biskyrmions are stable above a certain threshold
in the film thickness. In films of insufficient thickness, (e.g.,
with the number of the atomic layers $N_{z}=10$ for $D/J=0.001$,
stable biskyrmions exist only at $\beta=$\,PMA/DDI$\,=1$, when
the DDI nearly compensates the effect of the PMA. Since this condition
is impossible to satisfy in practice, we conclude that stable biskyrmions
do not exist in thin non-chiral magnetic films that contain just a
few atomic layers. For the monolayer, $N_{z}=1$, no biskyrmions at
all were found. 

In thicker films (with, e.g., $N_{z}\sim100$ for $D/J=0.001$ and
thus $\delta/a\lesssim N_{z}$ ) stable biskyrmions exist within a
finite range of $\beta$ below $\beta=1$. For stronger stabilizing
fields, smaller biskyrmions of sizes $\lambda\lesssim d\lesssim\delta$
and the shape close to that provided by the Belavin-Polyakov 2D exchange
model, Eqs.\ (\ref{biskyrmion_sx}) - (\ref{biskyrmion_sz}), have
been observed in our numerical studies. Their energy is close to $8\pi J$.
Such BP biskyrmions collapse at a critical value of the stabilizing
field. In the opposite limit, when the magnitude of the stabilizing
field decreases, the BP biskyrmion transforms into a bigger thin-wall
biskyrmion bubble. Regardless of the film thickness and other parameters
we did not find any stable biantiskyrmions with $d>0$.

The study presented here has focused on the stability of individual
biskyrmions. Biskyrmions observed in experiments \cite{Yu-2014,Zhang2016}
formed distorted triangular lattices. They were obtained from labyrinth
domains on increasing the magnetic field in a manner similar to how
lattices of $Q=1$ skyrmions have been observed. While our numerical
method easily generated lattices of $Q=1$ skyrmions in the films
with DMI, numerical generation of stable biskyrmion lattices in centrosymmetric
2D systems remains a challenging problem.

In our computations, stability of a single biskyrmion required an
external field while the biskyrmion lattice has been observed even
in a zero field \cite{Yu-2014}. This means that a sufficiently dense
biskyrmion lattice minimizes the sum of the exchange, DDI, and PMA
energies at $H=0$, similarly to what happens in the domain state.

For practical applications one has to be able to generate and manipulate
individual biskyrmions. It has been demonstrated that skyrmions can
be created, annihilated and moved by current-induced spin-orbit torques
\cite{Yu-NanoLet2016,Fert-Nature2017,Legrand-Nanolet2017}. Individual
$Q=1$ skyrmion bubbles have been generated by pushing elongated magnetic
domains through a constriction using an in-plane current \cite{Jiang-Sci2015,Hoffmann-PhysRep2017}.
These methods may not be suited for creating biskyrmions.

It has been shown that small skyrmions can be written and deleted
in a controlled fashion with local spin-polarized currents from a
scanning tunneling microscope \cite{Romming-Sci2013}. It has been
also demonstrated that light-induced heat pulses of different duration
and energy can write skyrmions in a magnetic film in a broad range
of temperatures and magnetic fields \cite{Berruto-PRL2018}. These
methods can be better suited for creating biskyrmions if experimentalists
find the way of using a scanning tunneling microscope with a double
tip or heat pulses of the shape resembling biskyrmions.

Recently, it has been experimentally demonstrated and confirmed through
micromagnetic computations that stripe domains in a film can be cut
into $Q=1$ skyrmions by the magnetic field of the tip of a scanning
magnetic force microscope (MFM) \cite{Senfu-APL2018}. Writing individual
$Q=1$ skyrmions by the MFM tip have been studied theoretically in
Ref.\ \onlinecite{AP-2018}. One can also use for that purpose
magnetic nanoparticles of the kind used in nanocantilevers for mechanical
magnetometry \cite{deLoubens}.

A simple modification of the above method, tailored to biskyrmions,
may be developed by using a double MFM tip consisting of two single
tips in close proximity to each other, or a nanocantilever with two
magnetic nanoparticles next to each other. We have tested this numerically
in the same manner as is described in detail in Ref.\ \onlinecite{AP-2018}.
A biskyrmion created this way in the numerical experiment relaxes
to the equilibrium size and shape determined by the parameters of
the film.

Since biskyrmions carry magnetic moments similar to those of $Q=1$
skyrmions, they can be utilized in a similar way for data storage
and information processing. The lack of the rotational symmetry in
a biskymion makes its orientation another useful parameter in addition
to the magnetic moment. It can open other functionalities in manipulating
such information carriers as well. For example, the magnitude of the
force exerted on a biskyrmion by the spin polarized current would
depend on the direction of the current.

\section{Acknowledgements}

This work has been supported by the grant No. OSR-2016-CRG5-2977 from
King Abdullah University of Science and Technology.

\end{document}